\documentclass[aip,jmp,preprint]{revtex4-1}
\usepackage{epsfig}
\usepackage{amsbsy,latexsym}
\usepackage{amsmath}
\usepackage{amssymb, mathrsfs}
\usepackage[mathscr]{eucal}
\usepackage{bm}
\usepackage[colorlinks=true,linkcolor=blue]{hyperref}%
\expandafter\ifx\csname package@font\endcsname\relax\else
 \expandafter\expandafter
 \expandafter\usepackage
 \expandafter\expandafter
 \expandafter{\csname package@font\endcsname}%
\fi
\def\<{\langle}
\def\>{\rangle}
\def\Tr{\operatorname{Tr}}

\def\:{\hbox{\bf :}}

\def\conv#1{\mathscr{#1}}

\def\dag{\dagger}
\def\geq{\geqslant}
\def\leq{\leqslant}

\def\dim{\operatorname{dim}}

\def\qed{$\,\blacksquare$\par}
\def\kk{\rangle\!\rangle}
\def\bb{\langle\!\langle}

\def\Ker{\operatorname{Ker}}
\def\Supp{\operatorname{Supp}}

\def\esp#1{\mathcal{#1}}
\def\esop#1{\mathcal{#1}}
\def\bas#1{\mathbb{#1}}

\newcommand{\one}{{I}}
\newcommand{\Hs}{{\cal H}}

\newcommand{\ket}[1]{| #1 \rangle}
\newcommand{\bra}[1]{\langle #1 |}

\newcommand{\GQI}{GQI } 
\newcommand{\extest}{\Theta(\Hs_2,\Hs_1)}

\newtheorem{Def}{Definition}
\newtheorem{lemma}{Lemma}

\newtheorem{corollary}{Corollary}
\newtheorem{theorem}{Theorem}
\newtheorem{remark}{Remark}

\def\qed{$\blacksquare$}

\begin{document}
\title{Extremal quantum protocols}

  \author{Giacomo Mauro D'Ariano}
  \affiliation{QUIT group,  Dipartimento di Fisica ``A. Volta'', and INFN Sezione di Pavia, via Bassi
    6, 27100 Pavia, Italy.}
  \author{Paolo Perinotti}
  \affiliation{QUIT group,  Dipartimento di Fisica ``A. Volta'', via Bassi
    6, 27100 Pavia, Italy.}
  \author{Michal Sedl\'ak}
  \affiliation{QUIT group,  Dipartimento di Fisica ``A. Volta'', via Bassi
    6, 27100 Pavia, Italy.}
    \affiliation{Institute of Physics, Slovak Academy of Sciences, D\'ubravsk\'a cesta 9, 845 11 Bratislava, Slovakia}

  \date{ \today}
\begin{abstract}
  Generalized quantum instruments correspond to measurements where the
  input and output are either states or more generally quantum
  circuits. These measurements describe any quantum protocol including
  games, communications, and algorithms.  The set of generalized
  quantum instruments with a given input and output structure is a
  convex set.  Here we investigate the extremal points of this set for
  the case of finite dimensional quantum systems and generalized
  instruments with finitely many outcomes. We derive algebraic
  necessary and sufficient conditions for extremality.
\end{abstract}

\maketitle
\section{Introduction}


Experiments in quantum theory can be modeled through quantum networks
that provide the natural description of an arbitrary quantum
procedure, corresponding to a causal sequence of steps. The most basic
building blocks of quantum networks are state preparations, state
transformations (channels and state reductions) and measurements.
Provided we have a quantum network, we can isolate open sub-circuits,
whose connections constitute the whole network.  Any optimization
problem in quantum theory can be seen as the search for the most
suitable sub-circuit for a specified purpose. For example, for
discrimination of states we need to optimize a measurement, or for
discrimination of channels we need to optimize the network into which
the channel is inserted. Open sub-circuits provide a representation
for the most general quantum protocol, where the gates represent the
sequence of operations performed by the agent that is
communicating, computing or applying a strategy for a quantum game.
From a more abstract point of view any sub-circuit represents the most
general input-output map that can be achieved via a quantum circuit,
that is 
called generalized quantum instrument (GQI)\cite{comblong}.  GQIs then
provide the mathematical description for any quantum protocol
including games, communications, and algorithms.  It is possible to
uniquely associate \cite{architecture} a positive operator to any
deterministic GQI---corresponding to a sub-circuit that does not
provide outcomes---in the same way as a positive operator is
associated to any channel through the Choi-Jamio\l kowski
correspondence.  More generally, it is possible to associate a set of
positive operators to any GQI \cite{memoryeff} in such a way that each
operator corresponds to a possible measurement outcome and summarizes
the probabilistic input-output behavior of the GQI as a sub-circuit,
conditionally on the outcome.  The advantage of this description comes
from neglecting the implementation details that are irrelevant for the
input-output behavior of the GQI within a quantum network, like
arbitrary transformations on ancillary systems, etc.  The set of GQIs
with the same input and output types is convex, since a random choice
of two different GQIs provides a convex combination of the
corresponding two input-output maps. 
It is thus clear that the description of quantum maps through GQIs
\cite{comblong} 
in optimization problems is convenient for two reasons. The first one
is that this approach gets rid of many irrelevant parameters, and the
second one is that the optimization problems are reduced to convex
optimization on suitably defined convex sets. Applications of GQIs in
optimization problems can be found in
\cite{cloning,covnet,tomo,infodist,learning,memory,ucomp,zimandiscr}.
The theory of GQIs was alternatively introduced\cite{comblong} as a
theory of higher order quantum functions, spawning interest in the
investigation of more computational consequences of the properties of
GQIs\cite{watrous,gutoski,costa1}. A similar approach to general
affine functions on convex subsets of state spaces was recently
published\cite{jenco}, explicitly inspired to the concept of GQIs and
quantum combs (namely singleton GQIs).

As a special case of GQIs, we have the elementary examples of states,
channels, and POVMs. The analysis of the extremality conditions for
states is trivial, and can be found in any textbook of quantum theory.
Algebraic extremality conditions for channels were provided in Ref.
\cite{choi}, while the conditions for POVMs 
were derived  later 
in Refs.  \cite{storm,partha,dlp,pellonpaa,chriribella1}. Other special cases of GQIs
are quantum combs \cite{architecture}, corresponding to deterministic GQIs, or quantum testers \cite{memoryeff,comblong}, which are GQIs with outputs
that are probability distributions.
While all GQIs could be decomposed into states, channels, and measurements,
it is much more practical to consider the corresponding networks as a whole.

Optimization tasks in quantum
information processing can be rephrased in terms of optimization of a
certain \GQI with respect to some particular figure of merit, which
is often a convex function on the set of GQIs and the maximum is achieved on an extremal
point of this set. Moreover, also for those problems that resort to convex
optimization or minimax problems, numerical optimization is enhanced
by the possibility of generating arbitrary extremal elements. For this
purpose, having an algebraic characterization is a crucial step.

In the present paper we consider the convex sets of GQIs, and characterize their extremal points for the case of
finite dimensional quantum systems and the instruments that have finitely many outcomes.
As special cases we obtain the extremality conditions for POVMs, channels,testers or instruments.

The paper is organized as follows. In section \ref{sec:combs} we
introduce the theoretical framework we use to describe quantum networks. In section \ref{sec:extgqi} we formulate the necessary and sufficient
condition of extremality for \GQI. Sections \ref{sec:extest}, \ref{sec:excombs}, and \ref{sec:exinst} study the implications of the extremality
condition in the case of quantum testers, quantum channels, and quantum instruments, respectively. Finally, the summary of the results is placed in
section \ref{sec:conclusion}.

\section{Theory of quantum networks}
\label{sec:combs}
\begin{figure}
  \includegraphics[width=0.7\textwidth]{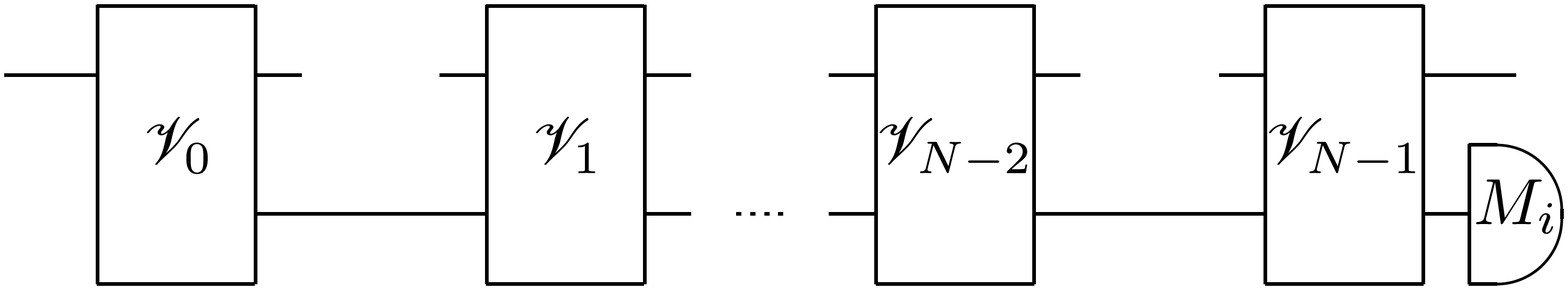}
  \caption{\label{f:gqisnet} The circuit implementation of a general
    GQI shows that GQIs correspond to the most general quantum
    network\cite{comblong}, and viceversa one can prove that any
    quantum circuit possibly involving measurements corresponds to a
    GQI. The transformations ${\conv V}_i$ are isometries, and $M_i$
    denotes a POVM. Notice that the measurement can always be
    postponed to the very last step.}
\end{figure}

Let us summarize some pieces of the theoretical framework of quantum networks introduced in \cite{comblong} that we will use.
An arbitrary quantum network $\mathcal{R}$ can be formally understood as a quantum memory channel \cite{architecture}, whose inputs and outputs are
labeled by even or odd numbers from $0$ to $2N-1$, respectively. The Hilbert spaces associated with these inputs and outputs can be in general
different and we denote them by $\Hs_i$ $i=0,\ldots,2N-1$. As it was shown already in \cite{architecture} deterministic quantum network $\mathcal{R}$
is fully characterized by its Choi-Jamiolkowski operator, i.e. a deterministic quantum $N$-comb $R$.
\begin{Def}
A deterministic quantum $N$-comb on $\Hs_0,\ldots,\Hs_{2N-1}$ is a positive operator $R\equiv R^{(N)}\in
\mathcal{L}(\Hs_0\otimes\cdots\otimes\Hs_{2N-1})$, which obeys the following normalization conditions
\begin{eqnarray}
\Tr_{2n-1}R^{(n)}&=&\one_{2n-2}\otimes R^{(n-1)} \quad 0\leq n\leq N \nonumber\\
\Tr_{1}R^{(1)}&=&\one_{0},
\label{normcascade}
\end{eqnarray}
where the operators $R^{(n)}$ are defined recursively.
\end{Def}
Positive operators $T\in \mathcal{L}(\Hs_0\otimes\cdots\otimes\Hs_{2N-1})$, such that $T\leq R$ for some deterministic quantum comb $R$, are called
non-deterministic quantum $N$-combs. An arbitrary probabilistic quantum network, whose different outcomes are indexed by $i=1,\ldots,M$ is described by a
collection of non-deterministic quantum $N$-combs $\{T_i\}_{i=1}^M$ defined as follows.
\begin{Def}
Generalized quantum $N$-instrument is a collection $\{T_i\}_{i=1}^M$ of non-deterministic quantum $N$-combs 
that sum up to a deterministic quantum comb 
\begin{eqnarray}
\sum_{i=1}^M T_i&=&R. \label{norminst}
\end{eqnarray}
\end{Def}
A realization theorem can be proved\cite{comblong}, providing the
interpretation of GQIs as the appropriate mathematical representation
for the most general quantum network, because any GQI can be
implemented through a quantum circuit as in Fig. \ref{f:gqisnet}, and
viceversa any quantum circuit possibly involving measurements
corresponds to a GQI.

For $M=1$ the corresponding network is deterministic and the set of generalized quantum instruments coincides with deterministic quantum combs. On the
other hand, if $N=1$ a generalized quantum instrument is a collection of completely positive maps forming a channel, which is usually called an
instrument. 
Another special case of generalized quantum instruments is provided by quantum testers.
\begin{Def}
A quantum $N$-tester is a generalized quantum $(N+1)$-instrument on $\Hs_0,\ldots,\Hs_{2N+1}$ with one-dimensional Hilbert spaces $\Hs_0$,
$\Hs_{2N+1}$.
\end{Def}
Quantum testers are analogous to the concept of Positive Operator Valued Measures (POVM) as they allow to express probability distributions for
arbitrary tests on quantum combs.

As we will show our analysis of the extremal points of the set of generalized quantum instruments provides necessary and sufficient conditions for
extremality, and leads to specific new conditions also for all the above mentioned special cases.

\section{Extremality condition for generalized quantum instruments}
\label{sec:extgqi}
In this section we shall apply the method of perturbations to find
extremal generalized quantum instruments. The perturbation method was
also used to determine extremal channels \cite{choi} and POVMs
\cite{dlp}. However, the application of the perturbation method to
GQIs does not come as a straightforward generalization of previous
results, because the richer structure of the normalization constraints
for GQIs requires a radically different analysis.

Let us consider arbitrary generalized quantum $N$-instrument $\{T_i\}_{i=1}^M$.
We denote by $\esp{V}_i$ the support of the operator $T_i$. The support of the sum of positive operators is the span of the supports of the summed
operators. Thus, the support of the normalization $R=\sum_{i=1}^M T_i$ is $\Hs_R\equiv \operatorname{Span}\{\esp{V}_i\}_{i=1}^{M}$. A set of operators
$\{ D_i\}_{i=1}^{M}$ is called a valid perturbation of \GQI $\{ T_i\}_{i=1}^{M}$ if and only if $\{ T_i\pm D_i\}_{i=1}^{M}$ are valid GQIs. Existence
of a perturbation has two major implications. First, the positivity of $T_i\pm D_i$ requires $D_i$ to be hermitian and to have support only in
$\esp{V}_i$. This is proved by the following lemma.

\begin{lemma}
Suppose that operators $T, D$ fulfill $T\geq0$, $D^\dag=D$. If $T\geq\pm D$, then $\Supp(D)\subseteq\Supp(T)$.
\end{lemma}

\begin{Proof}
The statement of the lemma can be equivalently formulated as
$\Ker(T)\subseteq\Ker(D)$. This can be proved considering the general
decomposition of a vector $\ket{\psi}$ as $\alpha\ket{\Psi_S}+\beta\ket{\Psi_K}$ where
$\Psi_S\in\Supp(T)$ and $\Psi_K\in\Ker(T)$. Then we have
\begin{eqnarray}
  |\alpha|^2\bra{\Psi_S}(T\pm D)\ket{\Psi_S} &\pm & 2\Re\alpha^{*} \beta\bra{\Psi_S}D\ket{\Psi_K}
  \pm |\beta|^2\bra{\Psi_K}D\ket{\Psi_K}\geq0,
\end{eqnarray}
for all $\alpha,\beta$. Choosing $\alpha=0$ one immediately obtains
$\bra{\Psi_K}D\ket{\Psi_K}=0$, which by the polarization identity implies also $\bra{\Psi^\prime_K}D\ket{\Psi_K}=0$ for all $\Psi^\prime_K\in\Ker(T)$. The
previous inequality can thus be rewritten
as follows
\begin{equation}
  |\alpha|^2\bra{\Psi_S}(T\pm D)\ket{\Psi_S}\pm2\Re\alpha^*\beta\bra{\Psi_S}D\ket{\Psi_K}\geq0,
\end{equation}
for all $\alpha,\beta$. Suitably choosing the phases of $\alpha$, 
$\beta$, one has
\begin{equation}
  |\alpha|^2\bra{\Psi_S}(T\pm D)\ket{\Psi_S}\mp 2|\alpha||\beta||\bra{\Psi_S}D\ket{\Psi_K}|\geq0,
\end{equation}
and for $\beta=\frac12$ and $|\alpha|>0$ we obtain
\begin{equation}
  |\alpha|\bra{\Psi_S}(T\pm D)\ket{\Psi_S}\geq|\bra{\Psi_S}D\ket{\Psi_K\>}|,
\end{equation}
for all $|\alpha|$, implying that $\bra{\Psi_S}D\ket{\Psi_K}=0$ holds for all $\Psi_S$ and $\Psi_K$. This together with
$\bra{\Psi^\prime_K}D\ket{\Psi_K}=0$ $\forall \Psi^\prime_K\in\Ker(T)$ allows us to conclude that $\bra{\psi}D\ket{\Psi_K}=0$ for every $\psi\in\Hs$,
i.e. $D\ket{\Psi_K}=0$ for all $\Psi_K$. This proves that $\Ker(T)\subseteq\Ker(D)$, or,
equivalently, $\Supp(D)\subseteq\Supp(T)$. \qed
\end{Proof}

As a consequence if we write operators $T_i$ in their spectral form $T_i = \sum_k \lambda^{(i)}_k \ket{v^{i}_k}\bra{v^{i}_k}$ then arbitrary hermitian
operator $D_i$ with support in $\esp{V}_i$ can be written as
\begin{eqnarray}
D_i = \sum_{n,m} D^{(i)}_{nm} \ket{v^{i}_n}\bra{v^{i}_m},
\end{eqnarray}
where $D^{(i)}_{nm}$ is a hermitian matrix with $r^2_i\equiv(\dim{\esp{V}_i})^2$ real parameters. We form a basis
$\bas{H}_i\equiv\{Q^{(i)}_j\}_{j=1}^{r^2_i}$ of hermitian operators with support in $\esp{V}_i$ and we define $\bas{D}_M:=\bigcup_{i=1}^M\bas{H}_i$.

The second consequence of requiring valid perturbed \GQI $\{T_i\pm D_i\}_{i=1}^M$ is that, due to the normalization condition (\ref{norminst}) the
perturbed \GQI has to sum up to deterministic $N$-combs $R_{\pm}$, which can be stated as
\begin{eqnarray}
\sum_{i=1}^M D_i = \Delta ,
\label{perturbcond1}
\end{eqnarray}
where $\Delta\equiv \pm (R_{\pm}-R)$ is an operator expressible as a difference of two deterministic quantum $N$-combs.
Using the parametrization of deterministic quantum combs developed in Appendix \ref{secparamcombs} it is clear that $\Delta$ lies in
$\esop{W}_C$, the subspace of operators spanned by the basis
\begin{eqnarray}
\bas{D}_{(N)}&\equiv&\{E^{(2N-1)}_i\otimes F^{(2N-2)}_j, \one_{2N-1,2N-2}\otimes E^{(2N-3)}_i \otimes F^{(2N-4)}_j, \ldots,
\one_{2N-1,\ldots,2}\otimes E^{(1)}_i \otimes F^{(0)}_j \} \nonumber
\end{eqnarray}
where $\{E^{(k)}_i\}_{i=2}^{d_k^2}$ is a basis of traceless hermitian operators on $\Hs_k$, and  $\{F^{(k)}_j\}_{j=1}^{d^2_k\ldots d^2_0}$ is basis of all
hermitian operators acting on $\Hs_k\otimes\Hs_{k-1}\otimes\cdots \otimes\Hs_0$. On the other hand, due to the positivity requirement for  $R_{\pm}=R\pm\Delta$, $\Delta$ must be a hermitian operator with support in $\Hs_R$. Let us call $\esop{W}_S$ the subspace of hermitian operators with support in $\Hs_R$. Thus, the allowed perturbations of the normalization lie in the intersection $\esop{W}_I\equiv \esop{W}_S \cap \esop{W}_C$. The
relation between non-existence of a valid perturbation and the requirements on the operators $D_1,\ldots,D_M,\Delta$ is expressed by the following
theorem.

\begin{theorem}
\label{th1}
A generalized quantum $N$-instrument $\{ T_i\}_{i=1}^{M}$ acting on $\Hs_{2N-1}\otimes\cdots\otimes\Hs_0$ is extremal if and only if $\bas{D}_M \cup
\bas{D}_{(N)}$ is an linearly independent set of operators.
\end{theorem}

\begin{Proof}
We are going to prove the theorem by showing the equivalence of the negated statements, i.e. \GQI is not extremal if and only if the basis $\bas{D}_M
\cup\bas{D}_{(N)}$ is linearly dependent. It is easy to show that if a \GQI is not extremal then the basis $\bas{D}_M \cup\bas{D}_{(N)}$ is linearly
dependent. If a point of a convex set is not extremal then there exists a bidirectional perturbation to it. Hence, there exists a set of operators
$D_i$ such that $\{ T_i\pm D_i\}_{i=1}^{M}$ is a valid \GQI. In particular, due to at least one operator $D_i$ being non-zero we have Eq.
(\ref{perturbcond1}), which after expanding the LHS in $\bas{D}_M$ and RHS in $\bas{D}_{(N)}$ proves the linear dependence of basis
$\bas{D}_M\cup\bas{D}_{(N)}$.

In order to prove the converse statement we are going to show that if the basis $\bas{D}_M \cup\bas{D}_{(N)}$ is linearly dependent then there exists a
valid perturbation of the considered \GQI and hence it is not extremal. Linear dependence of $\bas{D}_M \cup\bas{D}_{(N)}$ means there exists a
non-zero vector consisting of all coefficients $D^{(i)}_{nm}, s_k$ such that
\begin{eqnarray}
\sum_{i,n,m} D^{(i)}_{nm} \ket{v^{i}_n}\bra{v^{i}_m} + \sum_k s_k G_k = 0,
\label{prooflindep}
\end{eqnarray}
where $D^{(i)}_{nm}$ are for each $i$ hermitian matrices, $\ket{v^{i}_n}$ are eigenvectors of $T_i$ and $G_k$ are basis elements of $\bas{D}_{(N)}$.
Let us recall that the basis $\bas{D}_{(N)}$ is by construction linearly independent, so all $D^{(i)}_{nm}$ can not be zero simultaneously.
We rewrite the equation (\ref{prooflindep}) as:
\begin{eqnarray}
\sum_{i,n,m} D^{(i)}_{nm} \ket{v^{i}_n}\bra{v^{i}_m}=\sum_k -s_k G_k\equiv \Delta.
\label{prooflindep1}
\end{eqnarray}
For each $i$ the operators on the LHS of (\ref{prooflindep1}) have support in the subspace $\esp{V}_i$. All subspaces $\esp{V}_i$ are included in the
support of the normalization $R$. Thus, the operator on the LHS of (\ref{prooflindep1}) belongs to an operator subspace $\esop{W}_S$. Since the RHS of
(\ref{prooflindep1}) is from subspace $\esop{W}_C$ it is clear that $\sum_k s_k G_k\in \esop{W}_S\cap \esop{W}_C=\esop{W}_I$. This implies that for
suitably small $\varepsilon$ the operator $R\pm \varepsilon \Delta$ is positive as well as all operators $T_i\pm \varepsilon D_i$. Thus, we have found
a valid perturbation of the \GQI $\{T_i\}_{i=1}^M$ showing that it is not extremal, which concludes the proof. \qed
\end{Proof}

\section{Extremality of quantum testers}
\label{sec:extest}
In this section we focus our attention to quantum testers, which can be used to solve problems like discrimination of quantum channels, or optimization
of quantum oracle calling algorithms and others, because they describe achievable probability distributions for all possible experiments with given
resources. More precisely, we consider quantum $N$-testers and we try to identify the extremal points of this set. We start by the analysis of
$1$-testers, also called Process-POVMs \cite{ziman}. A $1$-tester with $M$ outcomes is defined by positive operators $\{ T_i\}_{i=1}^{M}$ acting on
$\Hs_2\otimes\Hs_1$, which satisfy the normalization condition
\begin{eqnarray}
\sum_{i=1}^M T_i = \one_2 \otimes \rho_1,
\label{norm1tester}
\end{eqnarray}
where $\rho$ is a state on $\Hs_1$\cite{note1}. As before we denote by
$\esp{V}_i$ the supports of operators $T_i$. Let us denote the support
of $\rho$ by $\Hs_\rho$ and by $r=\dim{\Hs_\rho}$ the rank of $\rho$.
The $1$-tester $\{ T_i\}_{i=1}^{M}$ on $\Hs_2\otimes\Hs_1$ can be
considered as a valid $1$-tester on $\Hs_2\otimes\Hs_1^\prime$ for
arbitrary $\Hs_1^\prime$ that includes $\Hs_\rho$ (e.g.
$\Hs_1^\prime=\Hs_\rho$).

\subsection{Extremality condition for $1$-testers}
\label{cond1test}
In the following we express the general extremality condition from Theorem \ref{th1} for $1$-testers and we propose a slightly different extremality
condition, which is easier to check. The set $\bas{D}_{(N)}$ from Theorem \ref{th1} is in this case formed by the operators $\{\one_2\otimes
E^{(1)}_i\}_{i=2}^{d_1^2}$, where $\{E^{(1)}_i\}_{i=2}^{d_1^2}$ is a basis of trace zero hermitian operators on $\Hs_1$.

\begin{corollary}
A quantum $1$-tester $\{ T_i\}_{i=1}^{M}$ is extremal if and only if there exists only a trivial solution of an equation
$\sum_{i=1}^M \sum_{n,m} D^{(i)}_{nm} \ket{v^{i}_n}\bra{v^{i}_m}+ \sum_{j=1}^{d_1^2-1} s_j \one_2\otimes E^{(1)}_j = 0,$
where $\forall i$ $D^{(i)}_{nm}$ are hermitian matrices and $s_j$ are real numbers.
\end{corollary}
Since the normalization of the perturbed tester must be supported inside the support of the original normalization, it is natural that,
$\mathbb{D}_{I}\equiv\{\one_2\otimes\sigma_l\}_{l=1}^{r^2 -1}$, the basis of trace zero operators supported under the original normalization
$\one_2\otimes\rho_1$ can be used in the Theorem \ref{th1} instead of $\bas{D}_{(N)}$.
\begin{theorem}
\label{th2}
A quantum $1$-tester $\{ T_i\}_{i=1}^{M}$ is extremal if and only if the equation
\begin{eqnarray}
\sum_{i=1}^M \sum_{n,m} D^{(i)}_{nm} \ket{v^{i}_n}\bra{v^{i}_m}+ \sum_{l=1}^{r^2-1} s_l \one_2\otimes\sigma_l = 0,
\label{perturbcond2}
\end{eqnarray}
where $D^{(i)}_{nm}$ are for each $i$ hermitian matrices and $s_l$ are real numbers, has only a trivial solution.
\end{theorem}

Actually, the basis $\mathbb{D}_{I}$ of the subspace $\esop{W}_I$ can be always used in the Theorem \ref{th1} and the proof still holds. However, for
$N\neq 1,2$ it is often easier to specify $\bas{D}_{(N)}$ rather than $\mathbb{D}_{I}$.

As we said for $1$-testers $\mathbb{D}_{I}$ is formed by trace zero operators supported under $\rho$ tensored with unity on $\Hs_2$ and this will help
us to get more insight to $1$-testers. The extremality condition for $1$-tester $\{ T_i\}_{i=1}^{M}$ from Theorem \ref{th2} allows us to give the
following bound
\begin{eqnarray}
\sum_{i=1}^M r_i^2 + r^2-1 \leq (r d_2)^2 ,
\label{bound1}
\end{eqnarray}
on the ranks $r_i$ of the operators $T_i$. The bound is derived by counting the number of elements of $\bas{D}_M \cup \bas{D}_{I}$  and realizing that
these operators should be linearly independent hermitian operators acting only on $\Hs_2\otimes\Hs_\rho$. From the bound (\ref{bound1}) it is clear
that the extremal tester can have the highest possible number of outcomes if $r=d_1$ and the ranks $r_i$ are as close to one as possible. Assuming all
$r_i$ are rank $1$ we get the bound on the number of elements of the extremal quantum $1$-tester
\begin{eqnarray}
M\leq d_1^2(d_2^2-1)+1 ,
\label{boundnumber}
\end{eqnarray}

\subsection{Classification of extremal $1$-testers}
\label{secclassif}
Let us now answer the question, which normalizations $\one\otimes\rho$ allow existence of extremal testers. For this purpose let us define a
superoperator $\xi_{\rho,U}$ that acts on linear operators on $\Hs_2\otimes\Hs_1$ as
\begin{eqnarray}
\xi_{\rho,U}(T_i)\equiv d_1 (\one\otimes\sqrt{\rho}\; U) \;T_i \; (\one\otimes U^\dagger \sqrt{\rho}).
\label{transftesters}
\end{eqnarray}
For any state $\rho$ with full rank (i.e. $r=d_1$) and any unitary $U$ acting on $\Hs_1$, the superoperator $\xi_{\rho,U}$ is invertible and preserves
positivity of operators. Using $\xi_{\rho,U}$ we can formulate the following theorem.

\begin{theorem}
\label{th3}
Suppose we have a full rank state $\rho$, a unitary operator $U$ and a $1$-tester $\{ T_i\}_{i=1}^{M}$ on $\Hs_2\otimes\Hs_1$ with $\sum_{i=1}^M
T_i=\one\otimes \frac{1}{d_1}\one$. Then the tester $\{ T^\prime_i\equiv \xi_{\rho,U}(T_i)\}_{i=1}^{M}$ on $\Hs_2\otimes\Hs_1$ has normalization
$\sum_{i=1}^M T^\prime_i=\one\otimes\rho$ and is extremal if and only if $1$-tester $\{ T_i\}_{i=1}^{M}$ is extremal.
\end{theorem}

\begin{Proof}
First, let us note that the form of $\xi_{\rho,U}$ guarantees positivity of $T^\prime_i$ and leads to the normalization
\begin{eqnarray}
\sum_{i=1}^M T_i^\prime=\xi_{\rho,U}(\one\otimes\frac{1}{d_1}\one)=\one\otimes\rho.
\end{eqnarray}
Now we prove that the tester $\{ T^\prime_i\}_{i=1}^{M}$ is extremal if the original tester $\{ T_i\}_{i=1}^{M}$ was. Let us stress that for any
extremal $1$-tester its normalization $\one\otimes\rho$ is (up to multiplication) the only operator of the form $\one\otimes X$ that is in the span of
the operators $D_i$. This holds, because the span of the operators $D_i\in\mathcal{L}(\esp{V}_i)\subseteq\mathcal{L}(\Hs_2\otimes\Hs_\rho)$ covers $\one\otimes\rho$ and
it is independent from $r^2-1$ dimensional subspace of traceless hermitian operators of the form $\one\otimes X$ due to linear independence
(\ref{perturbcond2}). Superoperator $\xi_{\rho,U}$ is invertible so it preserves linear independence. In our case this means that the basis
$\bas{H}^\prime_i$ of hermitian operators derived from $\xi_{\rho,U}(\ket{v^{i}_n}\bra{v^{i}_m})$ is linearly independent and spans the whole space of
hermitian operators that have support in the support of $T^\prime_i$. Moreover, due to extremality of the original tester $\{ T_i\}_{i=1}^{M}$
($\sum_{i=1}^M T_i=\one\otimes\frac{1}{d_1}\one$) and the invertibility of $\xi_{\rho,U}$ we can conclude that also  $\one\otimes\rho=\xi_{\rho,U}(\one\otimes\frac{1}{d_1}\one)$ is the only operator of the form $\one\otimes X$ that is in the span of $\bas{D}^\prime_M=\cup_{i=1}^M \bas{H}^\prime_i$.
Let us assume that the tester $\{ T^\prime_i\}_{i=1}^{M}$ is not extremal even though the original tester $\{ T_i\}_{i=1}^{M}$ was extremal. In other
words we assume that $\bas{D}^\prime_M$ is linearly dependent with traceless operators $\{\one_2\otimes\sigma_l\}_{l=1}^{r^2 -1}$. As a consequence
there must exist a traceless operator of the form $\one\otimes X$ in the span $D^\prime_i$. However, this is a contradiction, because the only operator
of such form is $\one\otimes\rho$ and has trace one. We conclude that the transformed tester $\{ T^\prime_i\}_{i=1}^{M}$ must be extremal.

In fact, the same argumentation can be used to prove that $\{T_i\}_{i=1}^{M}$ is extremal if $\{ T^\prime_i\}_{i=1}^{M}$ was, because $\xi_{\rho,U}$ is
invertible. Hence, for arbitrary extremal tester $\{ T^\prime_i\}_{i=1}^{M}$ with normalization $\one\otimes\rho$ using $(\xi_{\rho,U})^{-1}$ one
obtains extremal tester $\{ T_i\}_{i=1}^{M}$ with normalization $\one\otimes\frac{1}{d_1}\one$.
\qed
\end{Proof}

The theorem \ref{th3} is very useful, because to classify all extremal $1$-testers it suffices to classify extremal $1$-testers with normalization
$\one\otimes\frac{1}{d_1}\one$. More precisely, using $\xi^{-1}_{\rho,\one}$ each extremal tester is in one to one correspondence with an extremal
tester with normalization $\one\otimes\frac{1}{r}\Pi_\rho$, where $\Pi_\rho$ is a projector onto a support of $\rho$. This tester can be considered as
a tester on $\Hs_2\otimes\Hs_{\rho}$, where its normalization is of the above mentioned form $\one\otimes\frac{1}{d_1}\one$.
Thus, we can formulate the following corollary of theorem \ref{th3}.
\begin{corollary}
Extremal $1$-testers with $M$ outcomes exist either for all normalizations $\one\otimes\rho$ with given rank $r$ of $\rho$ or for none of them.
\end{corollary}

Let us now relate the set $\extest$ of extremal quantum testers with
normalization $\one\otimes\frac{1}{d_1}\one$ to the set
$\textit{P}(\Hs_2\otimes\Hs_1)$ of extremal POVMs on
$\Hs_2\otimes\Hs_1$. Namely, each extremal tester $\{ T_i\}_{i=1}^{M}$
with normalization $\one\otimes\frac{1}{d_1}\one$ defines an extremal
POVM $\{E_i=d_1 T_i\}_{i=1}^M$. This follows directly from the
extremality condition for quantum testers (\ref{perturbcond2}), which
necessarily requires the basis of hermitian operators with supports on
$\esp{V}_i$ to be linearly independent. This is exactly the necessary
and sufficient condition for the extremality of the POVM \cite{dlp}
$\{E_i\}_{i=1}^M$. Apart from the multiplicative difference in
normalization, we will prove later that extremal quantum testers with
normalization $\one\otimes\frac{1}{d_1}\one$ are a proper subset of
extremal POVMs on $\Hs_2\otimes\Hs_1$. On the other hand there are
extremal POVMs on $\Hs_2\otimes\Hs_1$, which cannot be rescaled to
form an extremal tester. One example are informationally complete
POVMs on $\Hs_2\otimes\Hs_1$ with $(d_1 d_2)^2$ outcomes.  Their
existence was proved in \cite{dlp} for any dimension, but they have
too many outcomes to form an extremal $1$-tester (see Eq.
(\ref{boundnumber})).

\subsection{Extremal $1$-testers with rank one normalization}
Having a tester with rank one normalization $\rho=\ket{\phi}\bra{\phi}$ implies that all the elements of the tester have the form $T_i=E_i\otimes\rho$,
where $E_i$ is positive operator acting on $\Hs_2$. Let us note that these testers correspond to preparation of a pure state $\rho$ and performing a
POVM $\{E^T_i\}_{i=1}^M$. Since the support of $\rho$ is one-dimensional, there are no traceless operators with support in $\Hs_\rho$. Thus, the
extremality condition (\ref{perturbcond2}) is in this case equivalent to linear independence requirement
\begin{eqnarray}
0&=& \sum_{i=1}^M \sum_{n,m} D^{(i)}_{nm} \;\ket{w^{i}_n}\ket{\phi}\bra{w^{i}_m}\bra{\phi}=
\left( \sum_{i=1}^M \sum_{n,m} D^{(i)}_{nm} \ket{w^{i}_n}\bra{w^{i}_m} \right) \otimes\ket{\phi}\bra{\phi} \;
\Rightarrow D^{(i)}_{nm}=0 \quad \forall i,n,m \nonumber
\end{eqnarray}
for the basis of hermitian operators on the supports of $E_i$. This is
precisely the necessary and sufficient condition of the extremality of
the POVM \cite{dlp} with elements $E_i$. Thus, the quantum tester
$\{T_i=\ket{\phi}\bra{\phi}\otimes E_i\}_{i=1}^{M}$ is extremal if and
only if POVM $\{E_i\}_{i=1}^M$ is extremal. In particular, the number
of outcomes of the extremal quantum tester in this case cannot exceed
$d_2^2$, which is the number given by the bound (\ref{boundnumber})
and by the maximal number of elements of an extremal POVM
\cite{dlp} as well. On the other hand, a single outcome extremal
POVM $\{E_1=\one\}$ leads to an extremal $1$-tester
$\{T_1=\one\otimes\rho \}$ for arbitrary pure state normalization
$\rho$.
\begin{remark}
Actually, the only extremal single outcome $1$-testers are those with pure state normalization.
\end{remark}

\subsection{Extremal qubit $1$-testers}
For qubit tester ($d_1=d_2=2$) the rank $r$ of the normalization $\rho$ can be either one or two. If $\rho$ is a pure state ($r=1$) then the previous
section tells us that such extremal testers are in one to one correspondence with the extremal qubit POVMs, which can have at most four outcomes.
Hence, to classify all extremal qubit testers (based on section \ref{secclassif}) it remains to investigate qubit testers with normalization
$\rho=\one\otimes\frac{1}{2}\one$. We will identify extremal testers with two outcomes. Then we discuss the case  $2< k \leq 13$
(see bound \ref{boundnumber}) and we propose some ways how to construct such testers

\subsubsection{Two outcome testers}
Considering the ranks $r_1$, $r_2$ of the two parts of the tester, there are only three possibilities compatible with bound (\ref{bound1}):
$i)$ $(r_1,r_2)=(1,3)$, $ii)$ $(r_1,r_2)=(2,2)$, $iii)$ $(r_1,r_2)=(2,3)$, where we assume without loss of generality that $r_1\leq r_2$. As we already
mentioned the supports of the tester operators $T_i$ necessarily have to obey conditions for extremal POVMs on $\Hs_2\otimes\Hs_1$. In particular,
operators $T_i$ cannot have intersecting supports (see corollary $3$ in ref \cite{dlp}). This rules out $(r_1,r_2)=(2,3)$ case.

Let us now consider the case $i)$ $(r_1,r_2)=(1,3)$. In this case $T_1$ necessarily equals $\frac{1}{2}$ projector onto a pure state, because otherwise
the rank of $T_2=\frac{1}{2}\one\otimes\one-T_1$ would not be three. Consequently, we can write the tester as
\begin{eqnarray}
T_1&=&\frac{1}{2}\ket{\phi_1}\bra{\phi_1}; \label{qtester13}\\
T_2&=&\frac{1}{2}(\one\otimes\one-\ket{\phi_1}\bra{\phi_1})=\frac{1}{2}\sum_{i=2}^4\ket{\phi_i}\bra{\phi_i}, \nonumber
\end{eqnarray}
where vectors $\ket{\phi_i}\; i=1,\ldots,4$ form an orthonormal basis of $\Hs_2\otimes\Hs_1$. As we show in the appendix \ref{appr13}, the only
two-outcome testers of the above form that are not extremal are those with $\ket{\phi_1}$ being a product state.
Looking on how the considered type of testers transforms under superoperator $\xi_{\rho,\one}$ from equation (\ref{transftesters}) one can easily
conclude that also for arbitrary rank two normalization $\rho$ the two outcome testers with $(r_1,r_2)=(1,3)$ are extremal if and only if
$\ket{\phi_1}$ is not a product state.

The case $ii)$ $(r_1,r_2)=(2,2)$ has some similarities to the previous one. Since $T_1, T_2$ are both rank two and their sum is $\frac12
\one\otimes\one$, then  they both must be equal to $\frac12 P_i$, where $P_i$ are orthogonal projectors. Consequently, we can write the tester as
\begin{eqnarray}
T_1&=&\frac{1}{2}P_1=\frac{1}{2}(\ket{\phi_1}\bra{\phi_1}+\ket{\phi_2}\bra{\phi_2}); \nonumber\\
T_2&=&\frac{1}{2}P_2=\frac{1}{2}(\ket{\phi_3}\bra{\phi_3}+\ket{\phi_4}\bra{\phi_4}); \label{qtester22}
\end{eqnarray}
where vectors $\ket{\phi_i}\; i=1,\ldots,4$ form an orthonormal basis of $\Hs_2\otimes\Hs_1$. As we show in the appendix \ref{appr22} this type of
tester is not extremal only if $P_1=\one\otimes\ket{v}\bra{v}$ for some $\ket{v}\in\Hs_1$ or if the states $\ket{\phi_1}$, $\ket{\phi_3}$ can be chosen
as $\ket{\phi_1}=\ket{w}\otimes\ket{v}$, $\ket{\phi_3}=\ket{w^{\perp}}\otimes\ket{v}$ for some states $\ket{w}\in\Hs_2$, $\ket{v}\in\Hs_1$.
For arbitrary rank two normalization $\rho$ the conditions on extremality of this type of tester are very similar, but with $P_1,P_2$ playing the role
of projectors onto the support of $T_1, T_2$.

\subsubsection{$M$-outcome testers}
The analysis of extremal qubit testers for more than two outcomes is
very involved. For this reason, we provide only some examples how one
can construct them. Extremal qubit $1$-testers with $3$ or $4$
outcomes can be easily obtained by taking the extremal $2$-outcome
tester from Eq. (\ref{qtester22}) and splitting either one or both its
parts into rank one operators. Obviously this operation reduces the
subspace achievable by linear combination of operators with support on
$T_i$, thus the linear independence with the operators
$\sigma_i\otimes\one$ remains untouched and the tester obtained in
this way is extremal. A different approach allows us to generate
examples of extremal testers with up to $M\leq 10$ as follows. Let us
consider an extremal $2$-outcome tester from Eq. (\ref{qtester13}) and
let us split its element $T_2$ into $T^{\prime}_2,\ldots,
T^{\prime}_M$ in such a way that $\{2 T^{\prime}_i\}_{i=2}^M$ is an
extremal POVM on the support of $T_2$. By setting $T^{\prime}_1=T_1$
we obtain an extremal tester $\{T^{\prime}_i\}_{i=1}^M$, because we
are only restricting the operator span of allowed perturbations of the
elements $T_i$ and perturbations of
$T_2^{\prime},\ldots\,T_M^{\prime}$ are independent by construction.
Finally, one can use the technique of Heinossari and Pellonp\"a\"a
\cite{heinosaari1} (see Proposition 4) to construct extremal qubit
$1$-testers with $4\leq M \leq 13$ rank $1$ elements. The construction
generates $M+1$ outcome tester from the $M$ outcome tester until the
linear independence of rank $1$ elements with the operators
$\sigma_i\otimes\one$ can be kept (i.e. $M\leq 13$).

\section{Extremality of quantum channels}
\label{sec:excombs}
The aim of this section is to show how our general criterion from
Theorem \ref{th1} in the case of channels ($N=1$, $M=1$) relates to
known conditions of extremality. For channels mapping from
$\mathcal{L}(\Hs_0)$ to $\mathcal{L}(\Hs_1)$, we have
$\bas{D}_{(N)}=\{\sigma_a\otimes\one, \sigma_a\otimes\sigma_b\}$,
where $\{\sigma_a\}_{a=2}^{d_1^2}$, $\{\sigma_b\}_{b=2}^{d_0^2}$ are
basis of trace zero hermitian operators on $\Hs_1$, $\Hs_0$,
respectively. Suppose we want to test whether a channel $\mathcal E$
with Choi-Jamiolkowski operator $E$ is extremal. If we take the
spectral decomposition of $E=\sum_m |K_m\kk \bb
K_m|$ then the eigenvectors $|K_m\kk$ correspond through
isomorphism \cite{dket} $|A\kk=A\otimes\one
|I\kk$ (here $|I\kk\equiv\sum_i
\ket{i}\otimes\ket{i}\in\Hs_0^{\otimes 2}$) to Kraus operators $K_m$
of a minimal Kraus representation of channel $\mathcal E$. The
well-known Choi extremality condition \cite{choi} writes
\begin{eqnarray}
  \sum_{m,n} \alpha_{mn} K_m^\dagger K_n=0 \quad\Leftrightarrow \alpha_{mn}=0\quad\forall m,n.
  \label{eq:choi}
\end{eqnarray}
On the other hand, according to our Theorem \ref{th1} the condition
for extremality of channel $\mathcal E$ is that
\begin{align}
  \sum_{m,n} &\alpha_{mn} |K_m\kk \bb K_n|+ \sum_a \beta_a \sigma_a\otimes\one + \sum_{a,b} \gamma_{ab} \sigma_a\otimes\sigma_b=0 ,\nonumber\\
  &\Leftrightarrow \alpha_{mn}=0,\ \forall m,n,\quad \beta_a=0\ \forall a,\quad\gamma_{ab}=0\ \forall a,b.
  \label{our}
\end{align}
We will now prove the following theorem
\begin{theorem}
  The conditions in Eq.~\eqref{eq:choi} and Eq.~\eqref{our} are equivalent.
\end{theorem}

\begin{Proof} In order to prove that the condition in
  Eq.~\eqref{eq:choi} implies the condition in Eq.~\eqref{our}, it is
  sufficient to suppose that Eq.~\eqref{our} holds, and to take the
  partial trace on the Hilbert space $\Hs_1$. We then get
  \begin{equation}
    \sum_{m,n=1}^{rank E} \alpha_{mn} K_m^T K_n^*=0\nonumber,
  \end{equation}
  which by condition Eq.~\eqref{eq:choi} implies $\alpha_{mn}=0$ for
  all $m,n$. Finally, by linear independence of $\{\sigma_a\otimes
  I,\sigma_a\otimes\sigma_b\}$, this also implies
  $\beta_a=0=\gamma_{ab}$ for all $a$ and $b$. Conversely, one can
  write $|K_m\kk \bb K_n|$ as
  \begin{equation}
    |K_m\kk \bb K_n|= \frac1{d_1}I_1 \otimes K^T_m K_n^* +\Delta_{mn},
    \label{partra}
  \end{equation}
  where $\Tr_1[\Delta_{mn}]=0$ for all $m,n$. This implies that
  $\Delta_{mn}$ belongs to the span of $\bas D_{(N)}$ for all $m,n$.
  Let us suppose that Choi's condition Eq.~\eqref{eq:choi} is not
  satisfied. Then there exist nontrivial coefficients $\zeta_{mn}$
  such that $\sum_{m,n}\zeta_{m,n}K^T_m K^*_n=0$. If we then take
  $\beta_a$, $\gamma_{ab}$ such that
  \begin{equation}
    \sum_{mn}\zeta_{mn}\Delta_{mn}=\sum_a\beta_a\sigma_a\otimes I+\sum_{ab}\gamma_{ab}\sigma_a\otimes \sigma_b,
  \end{equation}
  we have
  \begin{equation}
    \sum_{mn}\zeta_{mn}|K_m\kk\bb K_n|-\sum_a\beta_a\sigma_a\otimes I-\sum_{ab}\gamma_{ab}\sigma_a\otimes \sigma_b=0,
  \end{equation}
  in contradiction with Eq.~\eqref{our}.
\end{Proof}

\section{Extremality of quantum instruments}
\label{sec:exinst}
In contrast to a channel ($N=1$, $M=1$), which is specified by its
Choi-Jamiolkowski operator, an instrument ($N=1$, $M\geq 1$) is
characterized by a collection of Choi-Jamiolkowski operators
$\{N_i\}_{i=1}^M \subseteq \mathcal{L}(\Hs_1\otimes\Hs_0)$, which sum up to
Choi-Jamiolkowski operator of some channel $R$. The set
$\bas{D}_{(N)}=\{\sigma_a\otimes\one, \sigma_a\otimes\sigma_b\}$ from
Theorem \ref{th1} is the same as for channels, because it depends only
on $N$, the number of teeth of GQI, but not on $M$ the number of
outcomes of the instrument. We can take the spectral decompositions of
all the Choi-Jamiolkowski operators of the instrument $N_i=\sum_m
|K^{(i)}_m\kk \bb K^{(i)}_m|$ and we can write the necessary and
suffiecient condition of extremality as follows.

\begin{corollary}
Instrument $\{N_i\}_{i=1}^M \subseteq \mathcal{L}(\Hs_1\otimes\Hs_0)$ is extremal if and only if equation
\begin{eqnarray}
\sum_{i,m,n}\alpha^i_{mn} |K^{(i)}_m\kk \bb K^{(i)}_n|+ \sum_a \beta_a \sigma_a\otimes\one 
+\sum_{a,b} \gamma_{ab} \sigma_a\otimes\sigma_b&=&0   \label{excondinst}
\end{eqnarray}
cannot be satisfied for non-trivial coefficients $\alpha^i_{mn},
\beta_a, \gamma_{ab}$.
\end{corollary}

Counting the terms in Eq. (\ref{excondinst}) that have to be linearly
independent elements of $\mathcal{L}(\Hs_1\otimes\Hs_0)$, we can
obtain a simple restriction on the ranks of the elements of the
extremal instrument.

\begin{corollary}
  An extremal instrument $\{N_i\}_{i=1}^M \subseteq
  \mathcal{L}(\Hs_1\otimes\Hs_0)$ satisfies the following inequality
\begin{eqnarray}
  \sum_{i} r_i^2 \leq (d_0)^2,
\end{eqnarray}
where $r_i$ denotes the rank of $N_i$ and $d_0=\dim \Hs_0$.
\end{corollary}

We will now prove a theorem that provides an equivalent, but more
practical, extremality condition for quantum instruments.

\begin{theorem}
  An instrument $\{\mathcal N_i\}_{i=1}^M$ with Choi-Jamio\l kowski
  operators $\{N_i=\sum_{m}|K_m^{(i)}\kk\bb K^{(i)}_m|\}_{i=1}^M$ is
  extremal if and only if the operators $\{K_m^{(i)\dagger}
  K^{(i)}_n\}$ are linearly independent.\label{theoinst}
\end{theorem}

\begin{Proof}
  Suppose that the operators $\{K_m^{(i)\dagger} K^{(i)}_n\}$ are
  linearly independent. Then if Eq.~\eqref{excondinst} is satisfied,
  also its partial trace over space $\Hs_1$ is satisfied, namely
  \begin{equation}
    \sum_{i,m,n}\alpha^i_{mn} {K^{(i)}_m}^T{K^{(i)}_n}^*=0,
  \end{equation}
  which implies $\alpha^{(i)}_{mn}=0$ for all $i,m,n$ and consequently
  also $\beta_a=0$ for all $a$ and $\gamma_{ab}=0$ for all $a,b$.
  Conversely, consider the extremality condition in
  Eq.~\eqref{excondinst} along with the following generalization of
  Eq.~\eqref{partra}
  \begin{equation}
    |K_m^{(i)}\kk\bb K_n^{(i)}|=\frac1{d_1}I_1\otimes {K^{(i)}_m}^T{K_n^{(i)}}^*+\Delta^{(i)}_{mn},
  \end{equation}
  where the operators $\Delta^{(i)}_{mn}$ belong to the span of $\bas
  D_{(N)}$. If the operators $\{K_m^{(i)\dagger} K^{(i)}_n\}$ are not
  linearly independent, then there are non-trivial coefficients
  $\zeta^{(i)}_{mn}$ such that
  $\sum_{i,m,n}\zeta^{(i)}_{mn}{K_m^{(i)}}^T{K^{(i)}_n}^*=0$. Then,
  taking $\beta_a$ and $\gamma_{ab}$ such that
  \begin{equation}
    \sum_{i,m,n}\zeta^{(i)}_{mn}\Delta^{(i)}_{mn}=\sum_a\beta_a\sigma_a\otimes I+\sum_{ab}\gamma_{ab}\sigma_a\otimes \sigma_b,
  \end{equation}
  we have
  \begin{equation}
    \sum_{i,m,n}\zeta^{(i)}_{mn}|K^{(i)}_m\kk\bb K^{(i)}_n|-\sum_a\beta_a\sigma_a\otimes I-\sum_{ab}\gamma_{ab}\sigma_a\otimes \sigma_b=0,
  \end{equation}
  in contradiction with Eq.~\eqref{excondinst}.
\end{Proof}

\subsection{Extremality of Von Neuman-L\"uders instruments}

Let us now consider instruments of the following type
\begin{equation}
  \mathcal N_i(\rho)=\sqrt P_i\rho\sqrt P_i,
  \label{sqrtinst}
\end{equation}
where $P_i$ is a POVM. Then, by theorem \ref{theoinst}, the instrument
is extremal if and only if the POVM $\{P_i\}_{i=1}^M$ is linearly
independent.  Indeed, the set $\{K^{(i)\dagger}_mK^{(i)}_n\}$ in this
case is provided precisely by $\{P_i\}_{i=1}^M$.

In particular, von Neuman-L\"uders instruments are extremal. Indeed,
every such instrument $\{ N_i\}_{i=1}^d$ is of the form of
Eq.~\eqref{sqrtinst} with $P_i=\Pi_i$ where
$\Pi_i\Pi_j=\delta_{ij}\Pi_i$. Using the last constraint it is easy to
prove that if $X=\sum_i \alpha_i \Pi_i=0$ then
$\Pi_jX=\alpha_j\Pi_j=0$ and consequently $\alpha_j=0$.

Since there exist POVMs that are not extremal, but have linearly
independent elements, one can easily construct examples of extremal
instruments, with non-extremal POVMs. For example, this is the case
with $d=2$ and $P_1=1/2|0\>\<0|$, $P_2=1/2|0\>\<0|+|1\>\<1|$.

\section{Conclusions}
\label{sec:conclusion}
The aim of this paper was to characterize the extremal points of the set of generalized quantum instruments (GQIs). Our main result is represented by
theorem \ref{th1}, which links extremality of the considered \GQI with linear independence of a set of operators. An important special case of GQIs are
Quantum testers. For quantum $1$-testers we derived necessary and sufficient criterion of extremality that differs from the application of general
theorem \ref{th1} and can be tested more easily. As a consequence of the criterion, we obtained a bound (\ref{bound1}) on the ranks of elements of the
extremal $1$-tester. We showed that the subsets of extremal $1$-testers with a fixed normalization are isomorphic if they have the same rank of the
normalization. This implies that to classify all extremal $1$-testers it suffices to study extremal $1$-testers with a completely mixed normalization
($\one_2\otimes\rho_1=\frac1{d_1}I_{21}$). We completely characterized qubit $1$-testers with $1$ and $2$ outcomes and provided techniques to construct
extremal qubit testers with up to $13$ outcomes, which is the maximal number allowed by the bound (\ref{boundnumber}).

In section \ref{sec:excombs} we apply our extremality condition from
theorem \ref{th1} to channels.  The resulting condition is different
from the well known criterion of Choi \cite{choi}, even though we
prove it to be equivalent. The section \ref{sec:exinst} presents the
first characterization of the extremality of
instruments. 
In particular, we show that instruments of the type defined in
Eq.~\eqref{sqrtinst} for POVMs $\{P_i\}_{i=1}^M$ with linearly independent
elements are extremal quantum instruments.

More generally, any quantum instrument determines not only a POVM,
when the quantum output is ignored, but also a quantum channel, when
the classical outcome is ignored. A natural question is then what
combinations of extremality can exist when we consider an instrument
along with the POVM and channel it defines. In appendix
\ref{app:examples} we present examples of instruments for seven out of
the eight possibilities. The question whether non-extremal instruments
exist, such that they determine extremal POVMs and extremal channels
is left as an open problem.

\section*{Acknowledgments}
We thank the anonymous referee for a question that stimulated us to write appendix \ref{app:examples}.
This work has been supported by the European Union through FP$7$ STREP project COQUIT and by the Italian Ministry of Education through grant PRIN 2008
Quantum Circuit Architecture.

\appendix

\section{Parametrization of the set of deterministic quantum combs}
\label{secparamcombs}
Suppose we want to choose such parametrization of the set of hermitian
operators in which the subset of deterministic quantum combs would
simply correspond to positive operators that have some parameters
fixed (e.g. to zero). Let us consider a quantum $N$-combs $R\in
\mathcal{L}(\Hs_{2N-1}\otimes\ldots\otimes\Hs_0)$. For each of the
Hilbert spaces $\Hs_k$ $k\in\{0,..,2N-1\}$ we choose a basis of
hermitian operators on $\Hs_k$ $\{E^{(k)}_a\}_{a=1}^{d_k^2}$ such that
$E^{(k)}_1=\one$ and all the other elements have zero trace. Taking
the tensor product of the basis elements 
for all the Hilbert spaces $\Hs_k$ we obtain
a basis of hermitian operators $\{ E^{(2N-1)}_{a_{2N-1}}\otimes\ldots\otimes E^{(0)}_{a_{0}} \}$
on $\Hs_{2N-1}\otimes\ldots\otimes\Hs_{0}$.

Let us now use this basis to illustrate the normalization cascade requirements on the quantum $1$-combs i.e. Choi operators of quantum channels. In
this case a quantum channel mapping from ${\mathcal L}(\Hs_0)$ to ${\mathcal L}(\Hs_1)$ is represented via Choi-Jamiolkowski isomorphism by a positive
operator $R\in \mathcal{L}(\Hs_1\otimes\Hs_0)$, which has to fulfil equation $\Tr_1{R}=\one_0$. Using our basis arbitrary $R$ can be written as
\begin{eqnarray}
R&=&\sum_{a_1=1}^{d^2_1}\sum_{a_0=1}^{d^2_0} c_{a_1 a_0}\, E^{(1)}_{a_1}\otimes E^{(0)}_{a_0} \nonumber\\
&=&c_{11}\one_1\otimes\one_0 +\one_1\otimes\sum_{a_0=2}^{d^2_1} c_{1 a_0}  E^{(0)}_{a_0}+ \label{chanpar1}\\
& &+\sum_{a_1=2}^{d^2_1}E^{(1)}_{a_1}\otimes\sum_{a_0=1}^{d^2_0} c_{a_1 a_0} E^{(0)}_{a_0} \nonumber
\end{eqnarray}
Let us now look how the three terms of the RHS of Eq. (\ref{chanpar1}) contribute to $\Tr_1(R)$. The first two terms do contribute, whereas the
remaining one does not. The requirement of $\Tr_1(R)=\one_0$ translates into the following equations $c_{11}=\frac{1}{d_1}, \; c_{1i}=0 \;\forall
i=2,\ldots,d_0$ for parameters $c_{a_1 a_0}$. Thus, each quantum $1$-comb (Choi operator of a channel) can be written as
\begin{eqnarray}
R=\frac{1}{d_1} \one_{10}+\sum_i E^{(1)}_i\otimes A_i, 
\label{param1}
\end{eqnarray}
where 
$A_i$ are arbitrary hermitian operators on $\Hs_0$.  
Previous statements can be easily generalized to the case of general quantum combs. We shall first illustrate the relation of expansions for $R^{(n)}$
and $R^{(n-1)}$ and then write the expansion of general quantum $N$-comb. In our basis $R^{(n)}$ can be written as:
\begin{eqnarray}
R^{(n)}&=&\sum_{a_{2n-1},a_{2n-2}} E^{(2n-1)}_{a_{2n-1}}\otimes E^{(2n-2)}_{a_{2n-2}}\otimes L^{a_{2n-1},a_{2n-2}}, \nonumber \\
&=&\one_{2n-1,2n-2}\otimes L^{1,1}+\one_{2n-1}\otimes\sum_{j=2}^{d^2_{2n-2}} E^{(2n-2)}_{j}\otimes L^{1,j}+ \nonumber\\
& &+\sum_{i=2}^{d^2_{2n-1}}E^{(2n-1)}_{i}\otimes\sum_{j=1}^{d^2_{2n-2}} E^{(2n-2)}_{j}\otimes L^{i,j} \label{relrntorn1}
\end{eqnarray}
where
\begin{eqnarray}
L^{a_{2n-1},a_{2n-2}}=\sum_{a_{2n-3} \cdots a_0} c_{a_{2n-1}\cdots a_0}\,E^{(2n-3)}_{a_{2n-3}}\otimes\cdots\otimes E^{(0)}_{a_0} \nonumber
\end{eqnarray}
and we expanded the two sums in the same way as in (\ref{chanpar1}).
The normalization cascade (\ref{normcascade}) requires\cite{note2}
that
\begin{eqnarray}
L^{1,1}&=&\frac{1}{d_{2n-1}}R_{(n-1)}\;,\quad L^{1,j}=0 \; \forall j
\end{eqnarray}
and the operator $\sum_{j=1}^{d^2_{2n-2}} E^{(2n-2)}_{j}\otimes L^{i,j}$ can be an arbitrary operator on $\Hs_{2n-2}\otimes\cdots\otimes\Hs_0$.
As a result
\begin{eqnarray}
R_{(n)}&=&\frac{1}{d_{2n-1}}\one_{2n-1,2n-2}\otimes R_{(n-1)}+\sum_{i=2}^{d^2_{2n-1}} E^{(2n-1)}_i\otimes B_i, \nonumber\\
\label{recursiveform}
\end{eqnarray}
where 
$B_i\in {\mathcal L}(\Hs_{2n-2}\otimes\cdots\otimes\Hs_0)$.
Using the above relation recursively we can write the parametrization of the general deterministic $N$-comb as
\begin{eqnarray}
R_{(N)}&=&\frac{1}{d_{2N-1}d_{2N-3}\ldots d_1}\one_{2N-1,\ldots,0}+ \nonumber \\
& &+\sum_{i=2} E^{(2N-1)}_i\otimes B^{(2N-2)}_i+ \nonumber \\
& &+\one_{2N-1,2N-2}\otimes \sum_{i=2} E^{(2N-3)}_i \otimes B^{(2N-4)}_i+ \ldots+ \nonumber \\
& &+ \one_{2N-1,\ldots,2}\otimes \sum_{i=2} E^{(1)}_i \otimes B^{(0)}_i,  \label{param2}
\end{eqnarray}
where $B^{(k)}_i$ are arbitrary hermitian operators acting on
$\Hs_k\otimes\Hs_{k-1}\otimes\cdots \otimes\Hs_0$. Let us denote the
basis for hermitian operators $B_i^{(k)}$ as
$\{F^{(k)}_j\}_{j=1}^{d^2_k\ldots d^2_0}$. Consequently, the basis
used for the variable part (i.e. all terms in (\ref{param2}) except
the first) of the quantum comb is $\{E^{(2N-1)}_i\otimes F^{(2N-2)}_j,
\one_{2N-1,2N-2}\otimes E^{(2N-3)}_i \otimes F^{(2N-4)}_j, \ldots,
\one_{2N-1,\ldots,2}\otimes E^{(1)}_i \otimes F^{(0)}_j \}$ and we
denote it as $\bas{D}_{(N)}$. The operator basis $\mathbb{R}_{(N)}$
sufficient to expand arbitrary deterministic comb is then formed by
$\{\one_{2N-1,\ldots,0}\}\cup \bas{D}_{(N)}$.

\section {Two outcome qubit $1$-testers}
Suppose we have a two-outcome qubit $1$-tester $\{T_i=\frac{1}{2}P_i\}_{i=1}^2$ with normalization $\one_2\otimes \frac12 \one_1$ and $P_i$ being
orthogonal projectors. Equivalently to Theorem \ref{th2} we can say that the two outcome $1$-tester is extremal if and only if $\esp{V}_\sigma \cap
\esp{V}_T = 0$, where
$\esp{V}_\sigma=\operatorname{Span} \{\one\otimes\sigma_k\}_{k=x,y,z}$ and
$\esp{V}_T$ 
is the direct sum of the two subspaces of hermitian operators with support in $P_1$ and in $P_2$, respectively.
The non existence of the intersection of $\esp{V}_\sigma$ and $\esp{V}_T$ can be stated also as the impossibility to fulfill the following equation
\begin{eqnarray}
\sum_{i=1}^4 \lambda_i \ket{\phi_i}\bra{\phi_i}=\one\otimes(n_x\sigma_x + n_y\sigma_y + n_z\sigma_z),
\label{perturbc22}
\end{eqnarray}
where the left hand side of Eq. (B1) represents a generic element in $\esp{V}_T$ and the right hand side a generic element of $\esp{V}_\sigma$.
The set of $\{|\phi_i\>\}_{i=1}^4$ forms an orthonormal basis of vectors
belonging to $\Supp(P_1)\cup\Supp(P_2)$, and without loss of generality we can take $n^2_x + n^2_y + n^2_z = 1$. This guarantees that the RHS has the spectral decomposition of the following form
\begin{eqnarray}
\one\otimes\ket{v}\bra{v}-\one\otimes\ket{v^{\perp}}\bra{v^{\perp}}\nonumber
\end{eqnarray}
with two $+1$ eigenvalues and two $-1$ eigenvalues and vector $\ket{v}$ that can be arbitrary thanks to freedom in $n_x,n_y,n_z$.
Moreover projectors $P_i$ can be written as $P_1=\sum_{i=1}^{r_1}\ket{\phi_i}\bra{\phi_i}$ and $P_2=\sum_{i=r_1+1}^4\ket{\phi_i}\bra{\phi_i}$.
Let us now investigate the circumstances under which the equation can be fullfilled i.e. the tester is not extremal.

\subsection{Case $(r_1,r_2)=(1,3)$}
\label{appr13}
This type of tester must have the form
$\{T_1=\frac{1}{2}\ket{\phi_1}\bra{\phi_1},T_2=\frac{1}{2}\sum_{i=2}^4\ket{\phi_i}\bra{\phi_i}\}$.
The LHS of Eq. (\ref{perturbc22}) must have the same eigenvalues as
the RHS. Without loss of generality we can assume
$\lambda_1=\lambda_2=-\lambda_3=-\lambda_4$, because we can suitably
relabel $\ket{\phi_2},\ket{\phi_3},\ket{\phi_4}$. Hence, we have
\begin{eqnarray}
\ket{\phi_1}\bra{\phi_1}+ \ket{\phi_2}\bra{\phi_2}&=&\one\otimes \ket{e}\bra{e} \nonumber\\
\ket{\phi_3}\bra{\phi_3}+ \ket{\phi_4}\bra{\phi_4}&=&\one\otimes \ket{e^{\perp}}\bra{e^{\perp}},
\label{perturb13}
\end{eqnarray}
where $e=v$ or $e=v^{\perp}$ depending on $\lambda_1=+1$ or $\lambda_1=-1$, respectively. In both cases Eq. (\ref{perturb13}) implies that the qubit
$1$-tester of the form $T_1=\frac{1}{2}\ket{\phi_1}\bra{\phi_1}$ $ T_2=\frac{1}{2}(\one\otimes\one-\ket{\phi_1}\bra{\phi_1})$ is not extremal if and
only if $\ket{\phi_1}=\ket{f}\otimes\ket{e}$ is a product vector.

\subsection{Case $(r_1,r_2)=(2,2)$}
\label{appr22}
In this case the tester has the form $\{T_1=\frac{1}{2}P_1=\frac{1}{2}(\ket{\phi_1}\bra{\phi_1}+\ket{\phi_2}\bra{\phi_2})$,
$T_2=\frac{1}{2}P_2=\frac{1}{2}(\ket{\phi_3}\bra{\phi_3}+\ket{\phi_4}\bra{\phi_4})$. In order to fulfill the equation (\ref{perturbc22}) two
$\lambda_i$'s must be equal to $+1$ and two to $-1$. Thus, $\lambda_1$, $\lambda_2$ have either same signs or different signs. If
$\lambda_1=\lambda_2=\pm 1$ then the equation (\ref{perturbc22}) can be fulfilled if and only if $P_1=\one\otimes\ket{e}\bra{e}$, where $e=v$ for
$\lambda_{1,2}=1$ or $e=v^{\perp}$ for $\lambda_{1,2}=-1$. If $\lambda_1=-\lambda_2$ then we can assume without loss of generality that $\ket{\phi_3}$,
$\ket{\phi_4}$ are labeled so that $\lambda_1=\lambda_3$. We have
\begin{eqnarray}
\ket{\phi_1}\bra{\phi_1}+\ket{\phi_3}\bra{\phi_3}=\one\otimes\ket{e}\bra{e},
\label{pomeq1}
\end{eqnarray}
where $e=v$ or $e=v^\perp$ depending on $\lambda_1=1$ or $\lambda_1=-1$, respectively. This may hold only if $\ket{\phi_1}=\ket{f}\otimes\ket{e}$ and
$\ket{\phi_3}=\ket{f^\perp}\otimes\ket{e}$ for some vector $\ket{f}\in\Hs_2$. Due to equation (\ref{pomeq1})
$\ket{\phi_2}\bra{\phi_2}+\ket{\phi_4}\bra{\phi_4}=\one\otimes\ket{e^\perp}\bra{e^\perp}$ and $\ket{\phi_2}=\ket{h}\otimes\ket{e^\perp}$,
$\ket{\phi_4}=\ket{h^\perp}\otimes\ket{e^\perp}$ for some $\ket{h}\in\Hs_2$.
Thus, if $\lambda_1=-\lambda_2$ the tester is not extremal if and only if
\begin{eqnarray}
P_1&=&\ket{f}\bra{f}\otimes\ket{e}\bra{e}+\ket{h}\bra{h}\otimes\ket{e^\perp}\bra{e^\perp} \nonumber \\
P_2&=&\ket{f^\perp}\bra{f^\perp}\otimes\ket{e}\bra{e}+\ket{h^\perp}\bra{h^\perp}\otimes\ket{e^\perp}\bra{e^\perp}. \nonumber
\label{pomeq2}
\end{eqnarray}
for some $\ket{e}\in\Hs_1$, $\ket{f},\ket{h}\in\Hs_2$\cite{note3}. The
form of projectors $P_1$, $P_2$ can be equivalently stated as the
existence of a product vector $\ket{f}\otimes\ket{e}$ in the support
of $P_1$ such that $\ket{f^\perp}\otimes\ket{e}$ belongs to the
support of $P_2$.  From our derivation it should be clear that if
$P_1\neq \one\otimes\ket{e}\bra{e}$ for any $\ket{e}\in\Hs_1$ and
$P_1$ does not have the above mentioned form then $\{T_1,T_2\}$ is an
extremal qubit two-outcome tester.

\section{Extremality of an instrument and the POVM and the channel
derived from it}
\label{app:examples}
The present Appendix addresses the question about possible
combinations of extremality of an instrument and the POVM and channel
derived from it. We show feasibility of seven out of the eight
possible combinations, by providing an example for each of them. In
the following table, we enumearte the possible combinations, and we
define them by writing $+$ if the object in the corresponding column
(channel, POVM, instrument) is extremal, and $-$ otherwise.

\bigskip

\begin{tabular}{|c|c|c|c|}
\hline
Combination & Instrument & Channel & POVM \\
\hline
1 & + & + & + \\
2 & + & + & - \\
3 & + & - & + \\
4 & + & - & - \\
5 & - & + & + \\
6 & - & + & - \\
7 & - & - & + \\
8 & - & - & - \\
\hline
\end{tabular}

\bigskip

The existence of an instrument corresponding to combination number 5 is left as an
open problem. Here is a list of examples for each of the remaining
combinations.

\noindent {\bf Combination 1:}
The identical transformation is the most simple example of this kind. Also
constant mapping to a fixed pure state has the desired properties of
extremality.

\noindent {\bf Combination 2:}
Consider an instrument with two outcomes mapping a single qubit into
two-qubits. 
First, we define
\begin{eqnarray}
  P_0&:=&\frac13 \ket{0}\bra{0}+\frac23 \ket{1}\bra{1} \nonumber \\
  P_1&:=&\frac23 \ket{0}\bra{0}+\frac13 \ket{1}\bra{1} \nonumber \\
  W&:=&\ket{0}\bra{1}-\ket{1}\bra{0} \nonumber
\end{eqnarray}
and we define the Kraus operators of the instrument as follows
\begin{eqnarray}
  M_0&:=&\sqrt{P_0} \otimes \frac{1}{\sqrt{2}} (\ket{0} + \ket{1})  \nonumber \\
  M_1&:=&\frac{1}{\sqrt{2}} \sqrt{P_1} \otimes \ket{0} + \frac{1}{\sqrt{2}} W\sqrt{P_1} \otimes \ket{1},   \nonumber
\end{eqnarray}
where for each outcome $i=0,1$ we have only a single Kraus operator.
One can easily verify, that the induced POVM $\{P_0,P_1\}$ is not
extremal, but linear independence of its elements guarantees
extremality of the instrument. In order to check extremality of the
induced channel one needs to take the minimal Kraus representation and
check Choi's linear independence condition.

\noindent {\bf Combination 3:} The L\"uders instrument of a Von
Neumann measurement is an extremal instrument, which induces extremal
POVM and a non extremal channel.

\noindent {\bf Combination 4:} This desired type of instrument can be
constructed as in Eq.~\eqref{sqrtinst} with a POVM $\{P_i\}_{i=1}^M$, whose
elements are linearly independent and commute. As a simple example one
can take the qubit POVM $\{P_0, P_1\}$ defined in Combination 2.

\noindent {\bf Combination 6:} This type of instrument can be
constructed as follows. One takes an extremal channel, whose minimal
dilation has $N$ (more than one) Kraus operators $K_i$. Using these
operators we define two instruments with $N$ outcomes differing only
in the choice of Kraus operators that correspond to each outcome (e.g.
$M^{(1)}_i=K_i, M^{(2)}_i=K_{\sigma(i)}$, where $\sigma$ is a
permutation). Taking a convex combination of the two instruments
provides the desired example.

\noindent {\bf Combination 7:} This type of instrument can be
constructed as a convex combination of two instruments, which induce
the same POVM, but different channels.  One takes for example the
instrument $\{\mathcal N_i \}_{i=1}^M$ as in Eq.~\eqref{sqrtinst} for an
extremal POVM $\{P_i\}_{i=1}^M$, and mixes it with the same instrument, which
in addition applies an unitary channel $U$ on the quantum output.
Obviously, the induced channel differs, while the induced POVM remains
the same.

\noindent {\bf Combination 8:} For the construction of this example it
is sufficient to take a convex combination of two instruments, which
induce different POVMs, and different channels.


\begin{thebibliography}{99}
\bibitem{note1} Here the one dimensional Hilbert spaces $\Hs_3$,
  $\Hs_0$ are not considered, because
  $\Hs_3\otimes\Hs_2\otimes\Hs_1\otimes\Hs_0$ is isomorphic to
  $\Hs_2\otimes\Hs_1$.
\bibitem{note2} After performing the trace we are comparing the
  operators expanded in the same basis.
\bibitem{note3} Notice that for $\ket{f}=\ket{h}$ this form gives
  $P_1=\ket{f}\bra{f}\otimes\one$.  Moreover, the freedom in $\ket{e}$
  comes from $n_x,n_y,n_z$ and the freedom in $\ket{f},\ket{h}$
  relates to splitting of unity.
\end{thebibliography}
\end{document}